\documentclass[aps,preprint,nofootinbib]{revtex4}%
\usepackage{amsfonts}
\usepackage{amsmath}
\usepackage{amssymb}
\usepackage{graphicx}%
\providecommand{\U}[1]{\protect\rule{.1in}{.1in}}

\begin{document}
\preprint{ }
\title[Short title for running header]{Diffeomorphism Invariance in the Hamiltonian formulation of General Relativity}
\author{N. Kiriushcheva}
\email{nkiriush@uwo.ca}
\affiliation{Faculty of Arts and Social Science, Huron University College, N6G 1H3 and
Department of Applied Mathematics, University of Western Ontario, N6A 5B7,
London, Canada }
\author{S.V. Kuzmin}
\email{skuzmin@uwo.ca}
\affiliation{Faculty of Arts and Social Science, Huron University College, N6G 1H3 and
Department of Applied Mathematics, University of Western Ontario, N6A 5B7,
London, Canada }
\author{C. Racknor}
\email{cracknor@uwo.ca}
\affiliation{Physics and Astronomy Department and Department of Applied Mathematics,
University of Western Ontario, N6A 5B7, London, Canada }
\author{S.R. Valluri}
\email{valluri@uwo.ca}
\affiliation{Physics and Astronomy Department and Department of Applied Mathematics,
University of Western Ontario, N6A 5B7, London, Canada }
\keywords{General Relativity, Hamiltonian formulation }
\pacs{11.10.Ef}

\begin{abstract}
It is shown that when the Einstein-Hilbert Lagrangian is considered without
any non-covariant modifications or change of variables, its Hamiltonian
formulation leads to results consistent with principles of General Relativity.
The first-class constraints of such a Hamiltonian formulation, with the metric
tensor taken as a canonical variable, allow one to derive the generator of
gauge transformations, which directly leads to diffeomorphism invariance. The
given Hamiltonian formulation preserves general covariance of the
transformations derivable from it. This characteristic should be used as the
crucial consistency requirement that must be met by any Hamiltonian
formulation of General Relativity.

(Published in Phys. Lett. A 372 (2008) 5101)

\end{abstract}
\maketitle


\section{Introduction}

The Hamiltonian formulation of General Relativity (GR) is more than a half
century old. To cast GR into Hamiltonian form, a method which deals with
singular systems is required. A way of dealing with this problem, the
constrained dynamics, was proposed by Dirac in 1949 \cite{Dirac1950}. Dirac's
procedure was almost immediately applied by Pirani, Schild, and Skinner
\cite{Pirani} to GR using the covariant metric tensor as the canonical
variable.\ However, the analysis of \cite{Pirani} was not complete; the time
development of secondary constraints was not considered and, strictly
speaking, the closure of the Dirac procedure was not demonstrated. Neither the
Dirac conjecture that all first-class constraints generate gauge symmetries
\cite{Diracbook} nor explicit methods for derivation of gauge transformations
were known at that time.

A few years later, Dirac revisited this problem himself \cite{Dirac} with the
same choice of a canonical variable, the covariant metric tensor, and made a
modification of the Einstein-Hilbert (EH) Lagrangian to simplify the primary
constraints. This modification does not affect equations of motion but,
according to Dirac, \textquotedblleft can be achieved only at the expense of
abandoning four-dimensional symmetry\textquotedblright\ \cite{Dirac}. He
explicitly demonstrated that the canonical Hamiltonian is proportional to a
linear combination of secondary constraints. However, as in \cite{Pirani}, it
was not possible at that time to derive the gauge invariance which results
from the presence of first-class constraints.

Shortly after the appearance of Dirac's article \cite{Dirac}, a new set of
variables was introduced by Arnowitt, Deser, and Misner (ADM) (see \cite{ADM}
and references therein) and the geometrical interpretation of these new
variables was widely emphasized and discussed (e.g. see \cite{Kuchar}). \ 

The Dirac conjecture \cite{Diracbook} was converted into an algorithm
\cite{Castellani} (see also \cite{Henneaux, Banerjee}) long after the
appearance of \cite{Pirani, Dirac, ADM}. It was applied only to the most
widely accepted formulation based on the ADM variables. The complete
derivation of the gauge transformations from the full set of first-class
constraints in the ADM formulation actually appeared only recently
\cite{Saha}. However, a \textit{field-dependent} redefinition of the gauge
parameters found in \cite{Saha} is required to present the result in the
\textit{form} of diffeomorphism. These transformations \cite{Saha} have been
known for a long time \cite{Bergmann} and were partially derived in
\cite{Castellani} for the ADM formulation as an illustration of a general
procedure for the derivation of gauge transformations. In \cite{Bergmann} the
distinction between these transformations and the diffeomorphism
transformation was pointed out. In \cite{Saha} this distinction was called
\textquotedblleft the unity of the \textit{different}
symmetries\textquotedblright\ (our Italic).

Why is it in the case of GR, where the Lagrangian and the equations of motion
are invariant under the diffeomorphism transformation\footnote{In mathematical
literature the term \textit{diffeomorphism} refers to a mapping from one
manifold to another, which is differentiable, one-to-one, onto, with
differentiable inverse. However, in the literature on GR, the word
\textquotedblleft diffeomorphism\textquotedblright\ is often used as
equivalent of the transformation (\ref{eqn000}) (the semicolon
\textquotedblleft;\textquotedblright\ means a covariant derivative) and in our
article the latter meaning is employed. By a gauge invariance one usually
understands the invariance in the same coordinate frame of reference
\cite{Grishchuk, Popova} and with this respect the transformation
(\ref{eqn000}) serves as a gauge invariance of GR and can be written in
variety of forms (e.g. (\ref{eqnD2}) of Section V is more suitable for
comparison with the results there). Note that the right-hand side of
(\ref{eqn000}) is in fact a Lie derivative of the metric tensor \cite{Carmeli}
which is obviously a true tensor, so it immediately follows that it is not
affected by any change of coordinates (in other words, (\ref{eqn000}) is
generally covariant).}%

\begin{equation}
\delta_{(diff)}g_{\mu\nu}=-\xi_{\mu;\nu}-\xi_{\nu;\mu}~, \label{eqn000}%
\end{equation}
does the Hamiltonian formulation lead to \textquotedblleft
unity\textquotedblright, instead of giving the equivalent result? The
Lagrangian and Hamiltonian formulations are supposed to give the same results;
in particular, the same gauge invariance. This is well-known equivalence for
ordinary gauge field theories. In addition, we refer to the result of Samanta
\cite{Samanta} where (\ref{eqn000}) was derived using the Lagrangian method
\cite{Gitman-Tyutin} without introducing any new variables or modifying the EH Lagrangian.

The ADM change of variables, or rather the geometrical meaning related to it,
was criticized by Hawking on general grounds as contradictory to
\textquotedblleft the whole spirit of relativity\textquotedblright%
\ \cite{Hawking}. In our view, the disagreement between the results of
\cite{Saha} and \cite{Samanta} is a confirmation of Hawking's criticism. To
eliminate such a discrepancy, we reconsider the Hamiltonian formulation of GR
by neither using new variables (as was done by ADM) nor by making additional
modifications of the original action (as was done by Dirac). We revert to the
first canonical treatment of GR in \cite{Pirani} and prove by explicit
calculations that the Hamiltonian formulation of GR with the metric tensor as
the canonical variable leads to the symmetry of (\ref{eqn000}), as derived in
the Lagrangian approach of \cite{Samanta}. This is expected from a consistent
Hamiltonian formulation of GR.

\section{The Hamiltonian}

The starting point of the Hamiltonian formulation in \cite{Pirani, Dirac} is
the $\Gamma-\Gamma$ part \cite{Landau, Carmeli} of the Einstein-Hilbert
Lagrangian which is quadratic in first order derivatives of the metric tensor%

\begin{equation}
L=\sqrt{-g}g^{\alpha\beta}\left(  \Gamma_{\alpha\nu}^{\mu}\Gamma_{\beta\mu
}^{\nu}-\Gamma_{\alpha\beta}^{\nu}\Gamma_{\nu\mu}^{\mu}\right)  =\frac{1}%
{4}\sqrt{-g}B^{\alpha\beta\gamma\mu\nu\rho}g_{\alpha\beta,\gamma}g_{\mu
\nu,\rho} \label{eqn1}%
\end{equation}
where%

\[
B^{\alpha\beta\gamma\mu\nu\rho}=g^{\alpha\beta}g^{\gamma\rho}g^{\mu\nu
}-g^{\alpha\mu}g^{\beta\nu}g^{\gamma\rho}+2g^{\alpha\rho}g^{\beta\nu}%
g^{\gamma\mu}-2g^{\alpha\beta}g^{\gamma\mu}g^{\nu\rho}.
\]

To find momenta $\pi^{\alpha\beta}$ conjugate to $g_{\alpha\beta}$, we rewrite
(\ref{eqn1}) to explicitly separate terms with time derivatives of the
covariant metric tensor (the \textquotedblleft velocities\textquotedblright)%

\begin{equation}
L=\frac{1}{4}\sqrt{-g}B^{\alpha\beta0\mu\nu0}g_{\alpha\beta,0}g_{\mu\nu
,0}+\frac{1}{2}\sqrt{-g}B^{\left(  \alpha\beta0\mid\mu\nu k\right)  }%
g_{\alpha\beta,0}g_{\mu\nu,k}+\frac{1}{4}\sqrt{-g}B^{\alpha\beta k\mu\nu
t}g_{\alpha\beta,k}g_{\mu\nu,t} \label{eqn2.1}%
\end{equation}
where Latin alphabet is used for spatial components, $0$ for temporal one, and
$()$ or $\left(  ...\mid...\right)  $ -brackets\ indicate symmetrization in
two indices or two groups of indices, i.e.,%

\[
B^{\left(  \alpha\beta\right)  \gamma\mu\nu\rho}=\frac{1}{2}\left(
B^{\alpha\beta\gamma\mu\nu\rho}+B^{\beta\alpha\gamma\mu\nu\rho}\right)
,\left.  {}\right.  B^{\left(  \alpha\beta\gamma\mid\mu\nu\rho\right)  }%
=\frac{1}{2}\left(  B^{\alpha\beta\gamma\mu\nu\rho}+B^{\mu\nu\rho\alpha
\beta\gamma}\right)  .
\]

Using (\ref{eqn2.1}) we obtain%

\begin{equation}
\pi^{\gamma\sigma}=\frac{\delta L}{\delta g_{\gamma\sigma,0}}=\frac{1}{2}%
\sqrt{-g}B^{\left(  \left(  \gamma\sigma\right)  0\mid\mu\nu0\right)  }%
g_{\mu\nu,0}+\frac{1}{2}\sqrt{-g}B^{\left(  \left(  \gamma\sigma\right)
0\mid\mu\nu k\right)  }g_{\mu\nu,k}. \label{eqn3}%
\end{equation}

The explicit form of the first term is (see also \cite{Dirac})%
\begin{equation}
\frac{1}{2}\sqrt{-g}B^{\left(  \left(  \gamma\sigma\right)  0\mid\mu
\nu0\right)  }g_{\mu\nu,0}=\frac{1}{2}\sqrt{-g}g^{00}E^{\mu\nu\gamma\sigma
}g_{\mu\nu,0} \label{eqn3.1}%
\end{equation}
where%

\begin{equation}
E^{\mu\nu\gamma\sigma}=e^{\mu\nu}e^{\gamma\sigma}-e^{\mu\gamma}e^{\nu\sigma
},\left.  {}\right.  e^{\mu\nu}=g^{\mu\nu}-\frac{g^{0\mu}g^{0\nu}}{g^{00}}.
\label{eqn3.2}%
\end{equation}

Note that both $e^{\mu\nu}$ and $E^{\mu\nu\gamma\sigma}$are zero, unless all
$\mu$, $\nu$, $\gamma$ and $\sigma$ differ from $0$, and so from
(\ref{eqn3.1}) it immediately follows that we have $d$ primary constraints
($d$ is the dimension of spacetime)%

\begin{equation}
\phi^{0\sigma}=\pi^{0\sigma}-\frac{1}{2}\sqrt{-g}B^{\left(  \left(
0\sigma\right)  0\mid\mu\nu k\right)  }g_{\mu\nu,k}. \label{eqn4}%
\end{equation}
For $\gamma\sigma=pq$, equation (\ref{eqn3}) is invertible and giving%

\begin{equation}
g_{mn,0}=I_{mnpq}\frac{1}{g^{00}}\left(  \frac{2}{\sqrt{-g}}\pi^{pq}%
-B^{\left(  \left(  pq\right)  0\mid\mu\nu k\right)  }g_{\mu\nu,k}\right)
\label{eqn5}%
\end{equation}
where $I_{mnpq}$ is inverse of $E^{pqkl}$:%

\begin{equation}
I_{mnpq}=\frac{1}{d-2}g_{mn}g_{pq}-g_{mp}g_{nq},\left.  {}\right.
I_{mnpq}E^{pqkl}=\delta_{m}^{k}\delta_{n}^{l}. \label{eqn6}%
\end{equation}

The appearance of the singularity in (\ref{eqn6}) for $d=2$ corresponds to the
fact, that in two dimensions Eq. (\ref{eqn3}) cannot be solved for all
velocities. So the number of primary constraints equals the number of
independent components of the metric tensor.

The Hamiltonian is defined to be $H=\pi^{\alpha\beta}g_{\alpha\beta,0}-L$,
which after elimination of the velocities using (\ref{eqn5}) gives the total Hamiltonian%

\begin{equation}
H_{T}=H_{c}+g_{0\sigma,0}\phi^{0\sigma} \label{eqn6-2}%
\end{equation}
where the canonical Hamiltonian $H_{c}$ is%

\begin{align}
H_{c}  &  =\frac{1}{\sqrt{-g}g^{00}}I_{mnpq}\pi^{mn}\pi^{pq}-\frac{1}{g^{00}%
}I_{mnpq}\pi^{mn}B^{\left(  pq0\mid\mu\nu k\right)  }g_{\mu\nu,k}%
\label{eqn7a}\\
&  +\frac{1}{4}\sqrt{-g}\left[  \frac{1}{g^{00}}I_{mnpq}B^{\left(  \left(
mn\right)  0\mid\mu\nu k\right)  }B^{\left(  pq0\mid\alpha\beta t\right)
}-B^{\mu\nu k\alpha\beta t}\right]  g_{\mu\nu,k}g_{\alpha\beta,t}.\nonumber
\end{align}

\section{Closure of the Dirac procedure}

The fundamental PB are defined by%

\[
\left\{  g_{\alpha\beta},\pi^{\mu\nu}\right\}  =\frac{1}{2}\left(
\delta_{\alpha}^{\mu}\delta_{\beta}^{\nu}+\delta_{\alpha}^{\nu}\delta_{\beta
}^{\mu}\right)  \equiv\Delta_{\alpha\beta}^{\mu\nu}%
\]
and, following the standard procedure (see \cite{Diracbook, Gitman-Tyutin,
Kurt}), we have to calculate time development of primary constraints,
$\left\{  \phi^{0\sigma},H_{T}\right\}  $. Using (\ref{eqn4}) we now find the
PB among primary constraints%

\begin{equation}
\left\{  \phi^{0\sigma},\phi^{0\alpha}\right\}  =0 \label{eqn7b}%
\end{equation}
while, because $H_{c}$ is independent of $\pi^{0\alpha}$ (see (\ref{eqn7a})),
$\left\{  \phi^{0\sigma},H_{c}\right\}  $ cannot be proportional to primary
constraints. This leads to the secondary constraints:%

\begin{align}
\chi^{0\sigma}  &  =-\frac{1}{2}\frac{1}{\sqrt{-g}}\frac{g^{0\sigma}}{g^{00}%
}I_{mnpq}\pi^{mn}\pi^{pq}\label{eqn8}\\
&  +\frac{1}{2}\frac{g^{0\sigma}}{g^{00}}I_{mnpq}\pi^{mn}A^{\left(  pq0\mid
\mu\nu k\right)  }g_{\mu\nu,k}+\delta_{m}^{\sigma}\left[  \pi_{,k}%
^{mk}+\left(  \pi^{pk}e^{qm}-\frac{1}{2}\pi^{pq}e^{km}\right)  g_{pq,k}\right]
\nonumber\\
&  -\frac{1}{8}\sqrt{-g}\left(  \frac{g^{0\sigma}}{g^{00}}I_{mnpq}B^{\left(
\left(  mn\right)  0\mid\mu\nu k\right)  }B^{\left(  pq0\mid\alpha\beta
t\right)  }-g^{0\sigma}B^{\mu\nu k\alpha\beta t}\right)  g_{\mu\nu,k}%
g_{\alpha\beta,t}\nonumber\\
&  +\frac{1}{4}\sqrt{-g}\frac{1}{g^{00}}I_{mnpq}B^{\left(  \left(  mn\right)
0\mid\mu\nu k\right)  }g_{\mu\nu,k}g_{\alpha\beta,t}\left[  g^{\sigma
t}\left(  g^{00}g^{p\alpha}g^{q\beta}+g^{pq}g^{0\alpha}g^{0\beta}-2g^{\alpha
q}g^{0p}g^{0\beta}\right)  \right. \nonumber\\
&  -g^{\sigma p}\left(  2g^{00}g^{q\alpha}g^{t\beta}-g^{00}g^{\alpha\beta
}g^{qt}+g^{\alpha\beta}g^{0q}g^{0t}-2g^{q\alpha}g^{0\beta}g^{0t}-2g^{t\alpha
}g^{0\beta}g^{0q}+2g^{qt}g^{0\alpha}g^{0\beta}\right) \nonumber\\
&  \left.  +g^{0\sigma}\left(  2g^{\beta t}g^{\alpha p}g^{0q}-2g^{p\alpha
}g^{q\beta}g^{0t}-2g^{pq}g^{t\beta}g^{0\alpha}+2g^{pt}g^{q\beta}g^{0\alpha
}+g^{pq}g^{\alpha\beta}g^{0t}-g^{tp}g^{\alpha\beta}g^{0q}\right)  \right]
\nonumber\\
&  -\frac{1}{4}\sqrt{-g}g_{\mu\nu,k}g_{\alpha\beta,t}\left[  g^{\sigma
t}\left(  g^{\alpha\mu}g^{\beta\nu}g^{0k}+g^{\mu\nu}g^{\alpha t}g^{0\beta
}-2g^{\mu\alpha}g^{k\nu}g^{0\beta}\right)  \right. \nonumber\\
&  +g^{0\sigma}\left(  2g^{\alpha t}g^{\beta\mu}g^{\nu k}-3g^{t\mu}g^{\nu
k}g^{\alpha\beta}-2g^{\mu\alpha}g^{\nu\beta}g^{kt}+g^{\mu\nu}g^{kt}%
g^{\alpha\beta}+2g^{\mu t}g^{\nu\beta}g^{k\alpha}\right) \nonumber\\
&  \left.  +g^{\sigma\mu}\left(  \left(  g^{\alpha\beta}g^{\nu t}%
-2g^{\nu\alpha}g^{t\beta}\right)  g^{0k}+2\left(  g^{\beta\nu}g^{kt}-g^{\beta
k}g^{t\nu}\right)  g^{0\alpha}+\left(  2g^{k\beta}g^{\alpha t}-g^{\alpha\beta
}g^{kt}\right)  g^{0\nu}\right)  \right] \nonumber\\
&  -\frac{1}{2}\sqrt{-g}g^{00}E^{pqt\sigma}\left(  \frac{1}{g^{00}}%
I_{mnpq}B^{\left(  \left(  mn\right)  0\mid\mu\nu k\right)  }g_{\mu\nu
,k}\right)  _{,t}-\frac{1}{2}\sqrt{-g}B^{\left(  \left(  \sigma0\right)
k\mid\alpha\beta t\right)  }g_{\alpha\beta,kt}\nonumber
\end{align}

where%

\begin{equation}
A^{\alpha\beta0\mu\nu k}=B^{\left(  \alpha\beta0\mid\mu\nu k\right)  }%
-g^{0k}E^{\alpha\beta\mu\nu}+2g^{0\mu}E^{\alpha\beta k\nu}. \label{eqn9b}%
\end{equation}

We observe that the primary constraints (\ref{eqn4}) are the same as those
derived in \cite{Pirani} (see Eq. (6) of \cite{Pirani}) and provide the
correct limit of linearized GR \cite{GKK}, but our (\ref{eqn7a}) is different
from $H_{c}$ of \cite{Pirani} (see Eq. (8) of \cite{Pirani}). While Eq.
(\ref{eqn7a}) gives the correct limit of linearized GR, the $H_{c}$ of
\cite{Pirani} does not.

Of course, Eq. (\ref{eqn8}) can be presented in different forms, however this
form is convenient when performing an additional consistency check. Having
$H_{c}$ proportional to the secondary constraints is a common feature of all
generally covariant theories (e.g., see \cite{Gitman}) and this can be seen as
a good verification of the correctness of (\ref{eqn8}). Comparing
(\ref{eqn7a}) and (\ref{eqn8}) we obtain%

\begin{equation}
H_{c}=-2g_{0\sigma}\chi^{0\sigma}+\left(  2g_{0m}\pi^{mk}-\sqrt{-g}\left(
g_{0n}B^{\left(  \left(  nk\right)  0\mid\alpha\beta t\right)  }+g_{0\gamma
}B^{\left(  \left(  0\gamma\right)  k\mid\alpha\beta t\right)  }\right)
g_{\alpha\beta,t}\right)  _{,k}. \label{eqn10}%
\end{equation}

Let us continue to apply the Dirac procedure and consider the time derivatives
of secondary constraints, $\left\{  \chi^{0\sigma},H_{T}\right\}  $. We find that%

\begin{equation}
\left\{  \chi^{0\sigma},\phi^{0\gamma}\right\}  =-\frac{1}{2}g^{\sigma\gamma
}\chi^{00} \label{eqn112}%
\end{equation}
and%

\begin{align}
\left\{  \chi^{0\sigma},H_{c}\right\}   &  =-\frac{2}{\sqrt{-g}}I_{mnpq}%
\pi^{mn}\frac{g^{\sigma q}}{g^{00}}\chi^{0p}+\frac{1}{2}g^{\sigma k}%
g_{00,k}\chi^{00}+\delta_{0}^{\sigma}\chi_{,k}^{0k}\label{eqn122}\\
&  +\left(  -2\frac{1}{\sqrt{-g}}I_{mnpk}\pi^{mn}\frac{g^{\sigma p}}{g^{00}%
}+I_{mkpq}g_{\alpha\beta,t}\frac{g^{\sigma m}}{g^{00}}A^{\left(  pq\right)
0\alpha\beta t}\right)  \chi^{0k}\nonumber\\
&  -\left(  g^{0\sigma}g_{00,k}+2g^{n\sigma}g_{0n,k}+g^{n\sigma}\frac{g^{0m}%
}{g^{00}}\left(  g_{mn,k}+g_{km,n}-g_{kn,m}\right)  \right)  \chi
^{0k}.\nonumber
\end{align}

This completes the proof of closure of the Dirac procedure, i.e., no further
constraints arise. According to (\ref{eqn7b}), (\ref{eqn10}), (\ref{eqn112}),
and (\ref{eqn122}), all constraints are first-class. This is sufficient to
derive the gauge transformations. A number of algorithms of such a derivation
are available \cite{Henneaux, Banerjee}. We follow the approach that was
applied for the first time to field theories by Castellani \cite{Castellani}.

\section{The Generators}

The Castellani procedure \cite{Castellani} is based on a derivation of
generators of gauge transformations which are defined by \textit{chains} of
first-class constraints. One starts with primary first-class constraint(s),
$i=1,2,...$, and from them constructs chain(s) $\xi_{i}^{\left(  n\right)
}G_{\left(  n\right)  }^{i}$. Here $\xi_{i}^{\left(  n\right)  }$ is a time
derivative of the $i$th gauge parameter of order $n$ ($n=0,1,...$) and $n$
corresponds to the length of chain. The number of gauge parameters is equal to
the number of first-class primary constraints. Note, that the chains are
unambiguous when the primary constraints are defined. The functions
$G_{\left(  n\right)  }^{i}$ are calculated as follows:%

\begin{equation}
G_{\left(  1\right)  }^{\sigma}\left(  x\right)  =\phi^{0\sigma}\left(
x\right)  , \label{eqnC130}%
\end{equation}

\begin{equation}
G_{\left(  0\right)  }^{\sigma}\left(  x\right)  =-\left\{  \phi^{0\sigma
}\left(  x\right)  ,H_{T}\right\}  +\int\alpha_{\gamma}^{\sigma}\left(
x,y\right)  \phi^{0\gamma}\left(  y\right)  d^{3}y, \label{eqnC131}%
\end{equation}
where functions $\alpha_{\gamma}^{\sigma}\left(  x,y\right)  $ have to be
chosen so that the chain ends on \textit{the surface of primary constraints}%

\begin{equation}
\left\{  G_{\left(  0\right)  }^{\sigma},H_{T}\right\}  =primary
\label{eqnC140}%
\end{equation}
and the generator $G\left(  \xi_{\sigma}\right)  $ is given by%

\begin{equation}
G\left(  \xi_{\sigma}\right)  =\xi_{\sigma}G_{\left(  0\right)  }^{\sigma}%
+\xi_{\sigma,0}G_{\left(  1\right)  }^{\sigma}. \label{eqnC141}%
\end{equation}
To construct the generator (\ref{eqnC141}), we have to find functions
$\alpha_{\gamma}^{\sigma}\left(  x,y\right)  $ using the condition
(\ref{eqnC140})%

\begin{equation}
\left\{  G_{\left(  0\right)  }^{\sigma},H_{T}\right\}  =-\left\{
\chi^{0\sigma}\left(  x\right)  ,H_{T}\right\}  +\int\left\{  \alpha_{\gamma
}^{\sigma},H_{T}\right\}  \phi^{0\gamma}\left(  y\right)  d^{3}y+\int
\alpha_{\gamma}^{\sigma}\left\{  \phi^{0\gamma}\left(  y\right)
,H_{T}\right\}  d^{3}y. \label{eqnC145}%
\end{equation}

The second term of (\ref{eqnC145}) is irrelevant because it is zero on a
surface of primary constraints, according to the condition (\ref{eqnC140}).
The rest of the PBs being known (see (\ref{eqn112}) and (\ref{eqn122})) allows
one to just read off the functions $\alpha_{\gamma}^{\sigma}\left(
x,y\right)  $ from (\ref{eqnC145}) and obtain an explicit expression for
$G_{\left(  0\right)  }^{\sigma}$%

\begin{align}
G_{\left(  0\right)  }^{\sigma}\left(  x\right)   &  =-\chi^{0\sigma}-\left(
g_{00,0}\frac{1}{2}g^{0\sigma}+g_{0m,0}g^{\sigma m}-\frac{1}{2}g^{\sigma
k}g_{00,k}\right)  \phi^{00}-\delta_{0}^{\sigma}\phi_{,k}^{0k}\label{eqn160}\\
&  +\left(  -2\frac{1}{\sqrt{-g}}I_{mnpk}\pi^{mn}\frac{g^{\sigma p}}{g^{00}%
}+I_{mkpq}g_{\alpha\beta,t}\frac{g^{\sigma m}}{g^{00}}A^{\left(  pq\right)
0\alpha\beta t}\right)  \phi^{0k}\nonumber\\
&  -\left(  g^{0\sigma}g_{00,k}+2g^{n\sigma}g_{0n,k}+g^{n\sigma}\frac{g^{0m}%
}{g^{00}}\left(  g_{mn,k}+g_{km,n}-g_{kn,m}\right)  \right)  \phi
^{0k}.\nonumber
\end{align}

This completes the calculation of the generator (\ref{eqnC141}).

\section{Transformation of the Metric Tensor}

Transformations of fields can be found by calculating the following PB%

\begin{equation}
\delta g_{\mu\nu}=\left\{  G,g_{\mu\nu}\right\}  =\left\{  \xi_{\sigma
}G_{\left(  0\right)  }^{\sigma}+\xi_{\sigma,0}\phi^{0\sigma},g_{\mu\nu
}\right\}  . \label{eqn170}%
\end{equation}

Let us compare the result of this PB with (\ref{eqn000}) which we present in
equivalent but more suitable form for comparison with (\ref{eqn170})%

\begin{equation}
\delta_{(diff)}g_{\mu\nu}=-\xi_{\mu,\nu}-\xi_{\nu,\mu}+g^{\alpha\beta}\left(
g_{\mu\beta,\nu}+g_{\nu\beta,\mu}-g_{\mu\nu,\beta}\right)  \xi_{\alpha}.
\label{eqnD2}%
\end{equation}

For the time-time component, $g_{00}$, (\ref{eqn170}) gives%

\[
\delta g_{00}=\left\{  G,g_{00}\right\}  =-\frac{\delta}{\delta\pi^{00}%
}\left(  \xi_{\sigma}G_{\left(  0\right)  }^{\sigma}+\xi_{\sigma,0}%
\phi^{0\sigma}\right)
\]
so that non-zero contributions come only from those terms proportional to the
primary constraint $\phi^{00}$%

\begin{equation}
\delta g_{00}=\xi_{\sigma}\left(  g_{00,0}\frac{1}{2}g^{0\sigma}%
+g_{0k,0}g^{\sigma k}-\frac{1}{2}g^{\sigma k}g_{00,k}\right)  -\xi_{0,0}.
\label{eqnC166}%
\end{equation}
Putting $\mu=\nu=0$ and partially separating the space and time indices in Eq.
(\ref{eqnD2}), we have%

\[
\delta_{(diff)}g_{00}=-2\xi_{0,0}+g^{\sigma0}g_{00,0}\xi_{\sigma}+g^{\sigma
k}\left(  2g_{0k,0}-g_{00,k}\right)  \xi_{\sigma}%
\]
which is equivalent to (\ref{eqnC166}) up to a numerical factor%

\begin{equation}
2\left\{  g_{00},G\right\}  =\delta_{(diff)}g_{00} \label{eqnC168.1}%
\end{equation}
that can be incorporated into a parameter $\xi_{\sigma}\rightarrow2\xi
_{\sigma}$.

For the transformation of the space-time components, $g_{0k}$, we need the
corresponding part of the generator depending on $\phi^{0s}$%

\begin{align}
\delta g_{0k}  &  =-\frac{1}{2}\xi_{0,k}-\frac{1}{2}\xi_{k,0}-\frac{1}{2}%
\xi_{\sigma}\left(  -2\frac{1}{\sqrt{-g}}I_{mnpk}\pi^{mn}\frac{g^{\sigma p}%
}{g^{00}}+I_{mkpq}g_{\alpha\beta,t}\frac{g^{\sigma m}}{g^{00}}A^{\left(
pq\right)  0\alpha\beta t}\right) \label{eqn175}\\
&  +\frac{1}{2}\xi_{\sigma}\left(  g^{0\sigma}g_{00,k}+2g^{n\sigma}%
g_{0n,k}+g^{n\sigma}\frac{g^{0m}}{g^{00}}\left(  g_{mn,k}+g_{km,n}%
-g_{kn,m}\right)  \right)  .\nonumber
\end{align}

In the first bracket of (\ref{eqn175}) we substitute momenta in terms of the
metric (\ref{eqn3}) so that using (\ref{eqn9b}) we obtain%

\[
-\frac{1}{2}\xi_{\sigma}\frac{g^{\sigma m}}{g^{00}}\left(  -g^{0n}%
g_{mk,n}+g^{0\alpha}g_{\alpha k,m}+g^{0\alpha}g_{\alpha m,k}\right)
\]
and finally, after a simple rearrangement of terms in (\ref{eqn175}), we have%

\begin{equation}
\delta g_{0k}=-\frac{1}{2}\xi_{0,k}-\frac{1}{2}\xi_{k,0}+\frac{1}{2}%
\xi_{\sigma}g^{0\sigma}g_{00,k}+\frac{1}{2}\xi_{\sigma}g^{n\sigma}\left(
-g_{0k,n}+g_{0n,k}+g_{nk,0}\right)  .\label{eqnC180}%
\end{equation}
This is equivalent to the transformations of the corresponding components of
(\ref{eqnD2}) up to the same numerical factor that occurs in (\ref{eqnC168.1}).

For the transformation of the space-space components, $g_{nm}$, we need to
keep only terms in $G$ with $\pi^{pq}$ dependence, so that%

\[
\delta g_{nm}=\left\{  G,g_{nm}\right\}  =\frac{\delta}{\delta\pi^{nm}}\left(
\xi_{\sigma}\chi^{0\sigma}-\xi_{\sigma}\frac{2}{\sqrt{-g}}I_{abpk}\pi
^{ab}\frac{g^{\sigma p}}{g^{00}}\phi^{0k}\right)  .
\]

The second term produces contributions proportional to the primary constraints
and can be neglected on the surface of primary constraints. In $\chi^{0\sigma
}$ we need only momentum-dependent terms appearing in (\ref{eqn8}). We obtain then%

\begin{align*}
\delta g_{nm}  &  =-\frac{1}{2}\xi_{n,m}-\frac{1}{2}\xi_{m,n}+\xi_{\sigma
}\delta_{a}^{\sigma}\left(  e^{qa}\Delta_{nm}^{pk}-\frac{1}{2}\Delta_{nm}%
^{pq}e^{ka}\right)  g_{pq,k}\\
&  +\xi_{\sigma}\frac{g^{0\sigma}}{g^{00}}\frac{1}{2}\left(  -\frac{1}%
{\sqrt{-g}}I_{abnm}2\pi^{ab}+I_{nmcd}A^{\left(  cd\right)  0\mu\nu k}g_{\mu
\nu,k}\right)  .
\end{align*}
After substitution of $\pi^{ab}$ from (\ref{eqn3}) and using (\ref{eqn3.2}),
(\ref{eqn6}), and (\ref{eqn9b}), we have%

\begin{align}
\delta g_{nm}  &  =-\frac{1}{2}\xi_{n,m}-\frac{1}{2}\xi_{m,n}+\frac{1}{2}%
\xi_{0}\left(  g^{00}\left(  g_{0m,n}+g_{0n,m}-g_{nm,0}\right)  +g^{0p}\left(
g_{pm,n}+g_{pn,m}-g_{nm,p}\right)  \right) \label{eqnC191}\\
&  +\frac{1}{2}\xi_{p}\left(  g^{qp}\left(  g_{nq,m}+g_{mq,n}-g_{nm,q}\right)
+g^{0p}\left(  g_{0m,n}+g_{0n,m}-g_{nm,0}\right)  \right) \nonumber
\end{align}
which is again equivalent with (\ref{eqnD2}) up to the same \textit{numerical}
factor of $2$.

The transformations of the components of $g_{\mu\nu}$ in (\ref{eqnC166}),
(\ref{eqnC180}), (\ref{eqnC191}) can be combined into one covariant expression
(\ref{eqnD2}) or (\ref{eqn000}). This completes the proof of our original
statement that the Hamiltonian formulation gives the same result for the gauge
invariance of GR as the Lagrangian formulation \cite{Samanta}, as it should.
Moreover, this also demonstrates that we obtained the consistent Hamiltonian
formulation of GR because the transformation\ (\ref{eqn000}) derivable from it
is generally covariant.

\section{Conclusion}

We would like to point out some peculiarities of the Hamiltonian formulation
of GR. First of all, the \ transformation (\ref{eqnC191}) is valid on the
surface of primary constraints. In ordinary gauge field theories (e.g.,
Maxwell or Yang-Mills) this does not happen, because the right-hand side of
(\ref{eqnC140}) equals zero exactly. Another peculiarity of GR is the PB among
primary and secondary constraints is not zero by (\ref{eqn112}). Such
deviations from ordinary gauge theories actually can be expected. In ordinary
field theories, equations of motion are exactly invariant under a gauge
transformation, whereas the Einstein equations of motion, $G_{\mu\nu}%
=R_{\mu\nu}-\frac{1}{2}g_{\mu\nu}R=0$, transform into a combination of the
same equations \cite{Grishchuk}, i.e. are invariant only on-shell%

\begin{equation}
\delta_{\left(  diff\right)  }G_{\mu\nu}=-\xi^{\rho}G_{\mu\nu,\rho}-\xi_{,\mu
}^{\rho}G_{\nu\rho}-\xi_{,\nu}^{\rho}G_{\mu\rho}. \label{eqnGD}%
\end{equation}

In particular, because of this, the \textquotedblleft crucial
test\textquotedblright\ of having off-shell closure of the constraint algebra
\cite{Nicolai} seems as unreasonably strong for GR. Moreover, it is impossible
to expect off-shell closure for generators of a transformation which by itself
produces only on-shell invariance of the equations of motion (\ref{eqnGD}).

Going back to the \textquotedblleft spiritual\textquotedblright\ statement of
Hawking, we can make the conjecture or possibly draw the conclusion that if
diffeomorphism is derivable through the Lagrangian or Hamiltonian approach,
then the spirit of GR is \textquotedblleft alive\textquotedblright, as the
mathematical essence of the spirit of GR is to retain general covariance and
that diffeomorphism can be derived from its structure. Vice versa, if
diffeomorphism cannot be obtained in the Lagrangian or Hamiltonian
formulation, then such a formulation is not the equivalent to GR. In
particular, our Hamiltonian formulation of GR which does not resort to any
change of variables or modifications gives the same diffeomorphism invariance
as the Lagrangian approach \cite{Samanta}. On the contrary, the ADM
Hamiltonian formulation gives different transformations which can be presented
in the \textit{form} of diffeomorphism only after \textit{field-dependent} and
\textit{non-covariant} redefinition of parameters \cite{Saha}:%

\begin{equation}
\varepsilon_{ADM}^{\bot}=\left(  -g^{00}\right)  ^{-1/2}\xi_{diff}^{0},\left.
{}\right.  \left.  {}\right.  \varepsilon_{ADM}^{k}=\xi_{diff}^{k}%
-\frac{g^{0k}}{g^{00}}\xi_{diff}^{0}. \label{eqnBK}%
\end{equation}

Moreover, the equivalence of the Lagrangian and Hamiltonian formulations
implies that the ADM Lagrangian should give the same field-dependent
redefinition of gauge parameters (\ref{eqnBK}) if treated by the Lagrangian
method of \cite{Samanta}, and it is the expected result \cite{Samanta2}. The
equivalence of results in both formalisms for asymptotically flat space-times
was discussed in \cite{Petrov}.

In this paper we demonstrated that the consistent Hamiltonian formulation of
GR can be obtained by considering the metric tensor as the canonical variable
and not performing any change of variables. Following the Dirac conjecture and
applying the Castellani procedure, we derived the gauge generators from the
full set of first-class constraints. From these generators, without
redefinition of gauge parameters, we explicitly derived diffeomorphism
invariance in the Hamiltonian formulation of GR for the first time. In the
Hamiltonian formulation of GR, \ the covariance is not manifest but the final
result on gauge transformation is presented in manifestly covariant form as in
the Hamiltonian formulation of ordinary field theories. Our result is another
illustration of the resiliency of Einstein's General Relativity and any
approach, if correctly used, cannot violate its symmetry and leads to the
generally covariant transformation (\ref{eqn000}).

\textbf{Note added in proof}

After completion of this work it came to our attention that, by a different
from \cite{Banerjee} method which was used in \cite{Saha}, the
\textquotedblleft specific metric-dependent diffeomorphism" (\ref{eqnBK}) of
the ADM formulation was also obtained and extensively discussed in \cite{Pons}.

\section{Acknowledgements}

We would like to thank A.M. Frolov, D.G.C. McKeon, V.I. Rupasov and A.V.
Zvelindovsky for discussions and W.G. Cliff for inspiration and support. The
helpful correspondence with A.N. Petrov and S. Samanta is gratefully
acknowledged. The research was partially supported by the Huron University
College Faculty of Arts and Social Science Research Grant Fund.


\begin{thebibliography}{99}                                                                                               %


\bibitem {Dirac1950}P.A.M. Dirac, Can. J. Math. 2 (1950) 129.

\bibitem {Pirani}F.A.E. Pirani, A. Schild and R. Skinner, Phys. Rev. 87 (1952) 452.

\bibitem {Diracbook}P.A.M. Dirac, Lectures on Quantum Mechanics, Belfer
Graduate School of Sciences, Yeshiva University, New York, 1964.

\bibitem {Dirac}P.A.M. Dirac, Proc. Roy. Soc. A 246 (1958) 333.

\bibitem {ADM}R. Arnowitt, S. Deser and C.W. Misner, in: L. Witten (Ed.),
Gravitation: An Introduction to Current Research, Wiley, New York, 1962, p. 227.

\bibitem {Kuchar}K. Kuchar, in: W. Israel (Ed.), Relativity, Astrophysics and
Cosmology, Reidel Publ. Comp., Dordrecht, 1973, p. 237.

\bibitem {Castellani}L. Castellani, Ann. Phys. 143 (1982) 357.

\bibitem {Henneaux}M. Henneaux, C. Teitelboim and J. Zanalli, Nucl. Phys. B
332 (1990) 169.

\bibitem {Banerjee}R. Banerjee, H.J. Rothe and K.D. Rothe, Phys. Lett. B 479
(2000) 429; R. Banerjee, H.J. Rothe and K.D. Rothe, Phys. Lett. B 462 (1999) 248.

\bibitem {Saha}P. Mukherjee and A. Saha, arXiv: 0705.4358 [hep-th].

\bibitem {Bergmann}P.G. Bergmann and A. Komar, Int. J. Theor. Phys. 1 (1972) 15.

\bibitem {Grishchuk}L.P. Grishchuk, A.N. Petrov and A.D. Popova, Commun. Math.
Phys. 94 (1984) 379.

\bibitem {Popova}A.D. Popova and A.N. Petrov, Int. J. Mod. Phys. A 3 (1988) 2651.

\bibitem {Carmeli}M. Carmeli, Classical Fields: General Relativity and Gauge
Theory, John Wiley \& Sons, 1982.

\bibitem {Samanta}S. Samanta, arXiv: 0708.3300 [hep-th].

\bibitem {Gitman-Tyutin}D.M. Gitman and I.V. Tyutin, Quantization of Fields
with Constraints, Springer, Berlin, 1990.

\bibitem {Hawking}S.W. Hawking, in: S.W. Hawking and W. Israel (Eds.), General
Relativity. An Einstein Centenary Survey, Cambridge University Press,
Cambridge, 1979, p. 746.

\bibitem {Landau}L.D. Landau and E.M. Lifshitz, The Classical Theory of
Fields, fourth ed., Pergamon Press, Oxford, 1975.

\bibitem {Kurt}K. Sundermeyer, Constrained Dynamics, Lecture Notes in Physics,
vol. 169, Springer, Berlin, 1982.

\bibitem {GKK}K.R. Green, N. Kiriushcheva and S.V. Kuzmin, arXiv: 0710.1430 [gr-qc].

\bibitem {Gitman}G. F\"{u}l\"{o}p, D. M. Gitman and I. V. Tyutin, Int. J.
Theor. Phys. 38\textbf{ }(1999) 1941.

\bibitem {Nicolai}H. Nicolai, K. Peeters and M. Zamaklar, Class. Quantum Grav.
22 (2005) R193.

\bibitem {Samanta2}S. Samanta, private communication.

\bibitem {Petrov}A.N. Petrov, Int. J. Mod. Phys. D 4 (1995) 451; A.N. Petrov,
Int. J. Mod. Phys. D 6 (1997) 239.

\bibitem {Pons}J.M. Pons, D.C. Salisbury, L.C. Shepley, Phys. Rev. D 55 (19970
658; J.M. Pons, Class. Quantum Grav. 20 (2003) 3279. 


\end{thebibliography}
\end{document}